\documentclass{entcs} 

\usepackage{prentcsmacro}

\usepackage{epsfig}  
\usepackage{amssymb}
\usepackage{amsmath}
\usepackage{pstricks}

\newcommand{\one}{{\rm I}}
\newcommand{\two}{{\rm I}\hspace{-0.4mm}{\rm I}}
\newcommand{\four}{{\rm I}\hspace{-0.4mm}{\rm V}}
\def\endproof{\hfill$\Box$}

\def\lastname{Coecke and Edwards}

\begin{document}
\begin{frontmatter}
  \title{Toy Quantum Categories} 
  \author{Bob Coecke\thanksref{Bob}\thanksref{bob.coecke@comlab.ox.ac.uk}}
  \author{Bill Edwards\thanksref{Bill}\thanksref{bill.edwards@comlab.ox.ac.uk}}
  \address{Oxford University Computing Laboratory\\
    Wolfson Building, Parks Road, OX1 3QD Oxford, UK} 

    \thanks[Bob]{B.~C.~is supported by EPSRC Advanced Research Fellowship EP/D072786/1 entitled  \em The Structure of Quantum Information and its Ramifications for IT\em.}
    \thanks[bob.coecke@comlab.ox.ac.uk]{Email:
    \href{mailto:bob.coecke@comlab.ox.ac.uk} {\texttt{\normalshape
        bob.coecke@comlab.ox.ac.uk}}} 

    \thanks[Bill]{B.~E.~is supported by an EPSRC DTA Scholarship.}
    \thanks[bill.edwards@comlab.ox.ac.uk]{Email:
    \href{mailto:bill.edwards@comlab.ox.ac.uk} {\texttt{\normalshape
        bill.edwards@comlab.ox.ac.uk}}} 

\begin{abstract}
We show that Rob Spekken's toy quantum theory arises as an instance of our categorical approach to quantum axiomatics, as a (proper) subcategory of the dagger compact category ${\bf FRel}$ of finite sets and relations with the cartesian product as tensor, where observables correspond to dagger Frobenius algebras.  This in particular implies that the quantum-like properties of the toy model are in fact very general category-theoretic properties.  We also show the remarkable fact that we can already interpret complementary quantum observables on the two-element set in ${\bf FRel}$.
\end{abstract}
\end{frontmatter}   
\maketitle
Several authors have developed the idea that quantum mechanics (QM) can be expressed using categories rather than the traditional apparatus of Hilbert space \cite{AC,Baez}. Specifically,  \emph{dagger symmetric monoidal categories ($\dagger$-SMCs) with `enough' basis structures} are a suitable arena for describing many features of quantum mechanics \cite{CD,CP,CPP}. This is unsurprising since $\textbf{FdHilb}$, the category of finite dimensional Hilbert spaces, linear maps, which `hosts' standard (finite-dimensional) QM machinery, is such a category. Here, the basis structures correspond with orthonormal bases and `enough' means `there exist incompatible observables'. However, many features of quantum mechanics can be modelled in any category of this sort. 
This paper explores some concrete examples of `discrete' $\dagger$-SMCs with `enough' basis structures to model important QM features.  We demonstrate two important facts:
\begin{itemize}
\item Spekkens's toy model \cite{Spekkens} is an (interesting) instance of categorical quantum axiomatics;
\item Within the category ${\bf FRel}$ of finite sets, relations with the cartesian product as a tensor, the two element set $\{0,1\}$ comes equipped with two complementary observables\footnote{We became aware of this pair of complementary observables when discussing our axiomatisation of Spekkens's toy model with Jamie Vicary.}.
\end{itemize}
The first fact provides a very concise presentation of Spekkens's model and facilitates enriching it with additional quantum features; it also shows that its quantumness is not accidental but due to general abstract structural reasons.  The second fact is puzzling and even more obscures the the structural distinction between the quantum and the classical; moreover the category concerned provides a substantially  smaller/simpler quantum-like model than Spekkens's toy theory. 

Unlike $\textbf{FdHilb}$ these models will (of course) not be able to express quantum mechanics in its entirety, they will only exhibit certain features of QM. Why then are we so interested in them? Firstly, by examining what can and can't be done with these categories, we are better able to identify what mathematical structure is required, over and above the dagger symmetric monoidal and basis structure, to describe the full structure of QM. We can more clearly discern which mathematical features allow description of which physical features. Secondly, these discrete models have important computer science applications. They support \em model checking techniques \em to falsify statements about quantum mechanics.  Whenever a property is violated in one of these discrete models --verifying this can be automated-- we know that it cannot hold in standard QM either.

On a more technical note, recall that if a $\dagger$-SMC has dagger biproducts, then its finitary restriction --i.e.~restriction to objects of the form $\bigoplus_{i=1}^{i=n}{\rm I}$-- admits a matrix calculus with the endomorphisms of the involutive scalar monoid as entries \cite{AC}.  Basis structures then  arise in the obvious manner from this underlying matrix calculus.  Until now, these biproduct $\dagger$-SMCs provided all $\dagger$-SMCs with `enough' basis structures known to us.\footnote{Cobordism categories \cite{Baez} only admit a single basis on each object, hence not enough.}  This paper changes the situation.  The two models we describe manifestly take us out of the matricial realm:
\begin{itemize}
\item The category ${\bf Spek}$ is not a biproduct category.
\item While ${\bf FRel}$ is itself a biproduct category one of the two complementary observables on $\{0,1\}$ does not arise from biproduct structure; this contradicts claims made in \cite{AC} about ${\bf FRel}$.
\end{itemize}

\section{Preliminaries}\label{prelim}

A \emph{symmetric monoidal category} (SMC) is a category equipped with a bifunctor $-\otimes -$, an identity object ${\rm I}$, and various natural isomorphisms which obey several coherence conditions~\cite{MacLane}. Physically we imagine that the objects of the category represent systems and the morphisms represent processes which the systems undergo. The object $A\otimes B$ represents systems $A$ and $B$ viewed together as a single composite system. A \emph{dagger category} is equipped with an identity-on-objects involutive contravariant endofunctor $(-)^{\dagger}$. These structures are harmoniously integrated in a $\dagger$-SMC \cite{Selinger}. In $\textbf{FdHilb}$ the bifunctor is given by the tensor product, the identity object is the one-dimensional Hilbert space $\mathbb{C}$, and the dagger operation is the adjoint. 
States of a system $A$ are represented by morphisms of type $I\rightarrow A$ in analogy to the $\textbf{FdHilb}$ case where linear maps of type $\mathbb{C}\rightarrow \mathcal{H}$ are in bijective correspondence with elements of $\mathcal{H}$. 
We will asssume associativity and unit natural isosomorphisms to be strict \cite[p.257]{MacLane}.

A \emph{basis structure} \cite{CP} is an internal co-commutative comonoid 
\[
(A, \delta : A \rightarrow A \otimes A, \epsilon : A \rightarrow I) 
\]
which furthermore is \em isometric \em and  obeys the \em Frobenius identity\em, that is, respectively 
\begin{eqnarray}
\delta^{\dagger} \circ \delta = 1_{A} \qquad&\text{ and }&\qquad \delta \circ \delta^{\dagger} = (\delta^{\dagger}\otimes 1_{A}) \circ (1_{A}\otimes \delta^{\dagger})
\end{eqnarray}
Recently it has been proven that in $\textbf{FdHilb}$ there is a bijective correspondence between basis structures and orthonormal bases \cite{CPV} (hence the name). Concretely, given an orthonormal basis $\{|\psi_{i}\rangle\}_i$ of Hilbert space $\mathcal{H}$, $\delta$ \em copies \em the basis vectors while $\epsilon$ uniformly \em deletes \em them:
\begin{equation}\label{eq:copy}
\delta : \mathcal{H} \rightarrow \mathcal{H} \otimes \mathcal{H} :: |\psi_{i}\rangle \mapsto |\psi_{i}\rangle \otimes |\psi_{i}\rangle
\qquad\qquad\qquad
\epsilon : \mathcal{H} \rightarrow \mathbb{C} :: |\psi_{i}\rangle \mapsto 1
\end{equation}
As  in standard QM these abstract basis structures correspond to non-degenerate observables \cite{CP}.

Previous work by Abramsky and Coecke  \cite{AC} axiomatised QM using \emph{dagger compact closed categories}. In brief, for every object $A$ in such a category there exists a morphism $\eta_{A}: I \rightarrow A\otimes A$, obeying certain equations, which represents the maximally entangled state or Bell state. They showed that, for example, the teleportation protocol can be expressed using these morphisms. Such a morphism can always be formed from a basis structure as $\eta = \delta \circ \epsilon^{\dagger}$ \cite{CPP}. Thus, if all objects in a $\dagger$-SMC have basis structures then the category will be compact closed. Employing $\eta$ we can furthermore define abstract counterparts to the transpose 
\[
f^{\ast} := (1_{A}\otimes \eta^{\dagger}_{B})\circ(1_{A}\otimes f\otimes 1_{B})\circ(\eta_A \otimes 1_{B})
\]
and conjugate $f_{\ast}:=(f^{\dagger})^{\ast}$ of a map $f:A\rightarrow B$.

In general a quantum system will have many incompatible observables -- in $\textbf{FdHilb}$ these correspond to different bases and in the abstract categorical setting they will correspond to different basis structures. Of particular interest are observables which are \emph{complementary}. In a Hilbert space $\mathcal{H}$ these are defined as follows. Suppose observable $A$ has normalised eigenstates $|a_{i}\rangle$. A normalised state $|\psi\rangle$ is unbiased with respect to $A$ if for all $i$ we have $|\langle \psi | a_{i} \rangle |^{2} = 1/\text{dim}(\mathcal{H})$. Observable $B$ is complementary to $A$ if all of $B$'s eigenstates are unbiased with respect to $A$ and vice versa. Key examples of complementary observables are position and momentum, and $S_{x}$ and $S_{z}$ for a spin-1/2 system. 
For Hilbert spaces of finite dimension $n$ it has been shown that there can be no more than $n+1$ simultaneously complementary observables, and where $n$ is the power of a prime it has been shown that there are exactly $n+1$, although the general case is still an open problem. 

Coecke and Duncan recently gave an abstract characterisation of complementary basis structures~\cite{CD}. They begin by giving an abstract counterpart to the concept of an unbiased state. Given a basis structure $\{A,\delta,\epsilon\}$ and a state $\psi: I \rightarrow A$ we can form an operation $\Lambda^{\delta}(\psi) = \delta^{\dagger}\circ(\psi\otimes 1_{A})$.A state $\psi$ is \emph{unbiased} relative to $\{A,\delta,\epsilon\}$ if $\Lambda^{\delta}(\psi)$ is \emph{unitary} -- a morphism is unitary iff $f^{\dagger} = f^{-1}$. They show that the abstract notion of unbiased state coincides with the standard one in $\textbf{FdHilb}$. 
Then they inroduce abstract counterparts to basis vectors, that is, those vectors which are copied by $\delta$ in $\textbf{FdHilb}$ -- cf.~eq.(\ref{eq:copy}). A state $\phi:I\rightarrow A$ is \emph{classical} relative to $\{A,\delta,\epsilon\}$ if it is a \em real comonoid homomorphism\em,\footnote{In this paper we ignore non-trivial scalars, i.e.~morphisms of type $\one\to\one$, since they won't play a role.} that is,
\[
\delta\circ\phi = \phi\otimes\phi
\qquad\qquad \text{and} \qquad\qquad
(\epsilon\circ\phi)=1_{\rm I}\,.
\]

\begin{definition}\label{def:complementary}\cite{CD}
Two basis structures $\{A,\delta_{Z},\epsilon_{Z}\}$ and $\{A,\delta_{X},\epsilon_{X}\}$ are \emph{complementary} iff:
\parskip 0pt
\begin{itemize}
\item whenever $\phi:I\rightarrow A$ is classical for $\{A,\delta_{X},\epsilon_{X}\}$ it is unbiased for $\{A,\delta_{Z},\epsilon_{Z}\}$\,; 
\item whenever $\psi:I\rightarrow A$ is classical for $\{A,\delta_{Z},\epsilon_{Z}\}$ it is unbiased for $\{A,\delta_{X},\epsilon_{X}\}$\,; 
\item $\epsilon^{\dagger}_{X}$ is classical for $\{A,\delta_{Z},\epsilon_{Z}\}$ and $\epsilon^{\dagger}_{Z}$ is classical for $\{A,\delta_{X},\epsilon_{X}\}$.
\end{itemize}
\end{definition}

Clearly, in $\textbf{FdHilb}$ this definition coincides with the standard quantum mechanical  one \cite{CD}.   An an equivalent algebraic characterisation of complementary observables is provided by the following theorem.

\begin{theorem}\label{thm:complementary}{\rm\cite{CD}}
In a category with `enough points each  pair of complementary basis structures forms a (scaled) Hopf bialgebra with trivial antipode.
\end{theorem}

Coecke and Duncan go on to show that this abstract definition captures most of the behaviour of complementary observables in QM systems. 

\section{$\textbf{FRel}$}

Our first example is the category $\textbf{FRel}$ whose objects are finite sets and whose morphisms are relations. Viewed as a $\dagger$-SMC, the bifunctor $-\otimes -$ is the Cartesian product, and the identity object $\one$ is the single element set $\{\ast\}$. If $R$ is a morphism in $\textbf{FRel}$, i.e.~a relation between sets, then $R^{\dagger}$ is the relational converse.  

Every object has at least one basis structure -- which arises from the underlying biproduct structure. Let $N$ be a set with $n$ elements, which we will denote by $0,1,\dots, n-1$. The following two morphisms constitute a basis structure:
\begin{equation}
\delta : N \rightarrow N \times N :: i \sim (i,i) \qquad \text{and}\qquad\epsilon: N \rightarrow \{*\} :: i \sim \ast\,. 
\end{equation}
Denote the the two element set as  $\two$ and for convenience we set $\one:=\{\ast\}$.
On $\two$ this basis structure is $Z=\{\two,\delta_{Z},\epsilon_{Z}\}$ where:
\[
\delta_Z : \two\to \two\times \two::
\left\{\begin{array}{l}
0\sim (0,0)\\
1\sim (1,1)
\end{array}\right.
\qquad\qquad
\epsilon_Z:\two\to\one::
\left\{\begin{array}{l}
0\sim *\\
1\sim *
\end{array}\right.
\]
and has two classical points  and one unbiased point, namely, 
\begin{equation}
z_{0}: \one\rightarrow \two :: \ast \sim 0\quad ,\quad z_{1}: \one \rightarrow \two :: \ast \sim 1
\quad \text{and}\quad x_{0}: \one \rightarrow \two :: \ast \sim \{0,1\}\,.
\end{equation}
Our main observation in this section is that besides $Z$ the set $\two$ has another basis structure, namely $X=(\two,\delta_X,\epsilon_X)$\footnote{There is in fact a third basis structure obtained by exchanging the roles of $0$ and $1$ in the one above.  Since it has the same classical and unbiassed points as $(\two,\delta_X,\epsilon_X)$, it plays the same role, and hence we won't consider it explicitly here.} where:
\[
\delta_X : \two\to \two\times \two::
\left\{\begin{array}{l}
0\sim \{(0,0),(1,1)\}\\
1\sim \{(0,1),(1,0)\}
\end{array}\right.
\qquad\qquad
\epsilon_X:\two\to\one::
0 \sim *
\]
which now has $z_{0}$ and $z_{1}$ as unbiased points and $x_{0}$ as its single classical point.

\begin{theorem}
Basis structures $(\two,\delta_Z,\epsilon_Z)$ and $(\two,\delta_X,\epsilon_X)$ are complementary 
in the sense of both Def.~\ref{def:complementary} and Thm.~\ref{thm:complementary}.
\end{theorem}

Thus the two element set in ${\bf FRel}$ represents a system with two observables, complementary to one another, one with one classical state, the other with two. Compared to a standard qubit, which has a continuum of observables, each with two states, and for which at most three observables can be simultaneously complementary, it is clear that the two-element set is a far from perfect model of a qubit.  Yet it can still model a considerable amount of a qubit's behaviour. 

\begin{proposition}
The two-observable structure $(\two, (\delta_Z,\epsilon_Z), (\delta_X,\epsilon_X))$ in ${\bf FRel}$ is rich enough to simulate the quantum teleportation and dense coding protocols --including the classical communication and decoherence due to measurement.
\end{proposition}
\noindent{\bf Sketch of proof.} In \cite{CP} it was shown that quantum teleportation and dense coding can be simulated whenever we have a so called `Bell-basis' $(A, Bell: A\otimes A\to B)$ relative to a  basis structure $(B,\delta,\epsilon)$. 
One shows that given \em any \em pair of complementary basis structures $(A,\delta_Z,\epsilon_Z)$ and $(A,\delta_X,\epsilon_X)$, that
$\left(A\otimes A\,, \delta_{X\otimes Z}\,, \epsilon_{X\otimes Z}\right)$
with
\[
\delta_{X\otimes Z}=(1_A\otimes\sigma_{A,A} \otimes 1_A)\circ(\delta_X\otimes \delta_Z):A\otimes A\to(A\otimes A)\otimes(A\otimes A)
\]
and
\[
\epsilon_{X\otimes Z}= \epsilon_X\otimes \epsilon_Z:A\otimes A\to\one
\]
is also a basis structure, and, in particular, that 
\[
\left(A, (\delta_X^\dagger\otimes 1_A)\circ(1_A\otimes\delta_Z):A\otimes A\to A\otimes A\right)
\]
is always such a Bell-basis. Since $(\two,\delta_Z,\epsilon_Z)$ and $(\two,\delta_X,\epsilon_X)$ is a pair of complementary basis structures quantum teleportation and dense coding can be simulated with it.
\endproof\medskip\par

This ability to simulate full-blown teleportation  on $\two$ in ${\bf FRel}$ contradicts a claim made in the original categorical QM semantics paper  \cite{AC} --where only basis structure arising from biproduct structure was considered.  Any  other quantum phenomena shown in \cite{CD} to result from the existence  of complementary basis structures can be simulated on $\two$ in ${\bf FRel}$.

What is the origin of the `unexpected' basis structure $(\two,\delta_X,\epsilon_X)$? Morphisms of $\textbf{FdHilb}$, linear maps, can be represented as matrices of complex numbers. 
Morphisms of $\textbf{FRel}$, relations between sets, can also be represented by matrices with entries now drawn from the two element Boolean semiring $\mathbb{B}=(\{0,1\},\wedge,\vee)$. Suppose $R: X\rightarrow Y$ is a relation between sets $X$ 
and $Y$, whose members we list in some order. The $(i,j)^{\text{th}}$ element of the matrix 
representing $R$ is equal to $1$ if the $i^{\text{th}}$ element of $Y$ is related to the $j^{\text{th}}$ 
element of $X$, and equal to $0$ otherwise. Composition of two relations can be achieved by matrix 
multiplication, with Boolean $\lor$ and $\land$ taking the usual roles of addition and multiplication 
of scalars. 
In this matrix representation we have: 
\begin{equation}
\delta_{Z} = \left(\begin{array}{cc}
                              1&0\\
                              0&0\\
                              0&0\\
                              0&1
                              \end{array}\right)\ \ ,\ \ \text{    }
\epsilon_{Z} =  \left(\begin{array}{cc}
                                   1&1
                                   \end{array}\right)
\quad\text{and}\quad
\delta_{X} = \left(\begin{array}{cc}
                              1&0\\
                              0&1\\
                              0&1\\
                              1&0
                              \end{array}\right)\ \  ,\ \  \text{    }
\epsilon_{X} =  \left(\begin{array}{cc}
                                   1&0
                                   \end{array}\right)
\end{equation}
If 0 and 1 are instead interpreted as elements of $\mathbb{C}$, these matrices also 
represent basis structures in $\textbf{FdHilb}$. The basis structure labelled by $Z$ corresponds to copying the $\{|0\rangle, |1\rangle\}$-basis, while that labelled by $X$ corresponds to copying the $\{|+\rangle = \frac{1}{\sqrt{2}}(|0\rangle + |1\rangle), |-\rangle = \frac{1}{\sqrt{2}}(|0\rangle - |1\rangle)\}$-basis. To see this, observe that the prescriptions 
\[
\left\{\begin{array}{l}
|0\rangle\mapsto |00\rangle+|11\rangle\\
|1\rangle\mapsto |01\rangle+|10\rangle
\end{array}\right.
\qquad\qquad\text{and}\qquad\qquad
\left\{\begin{array}{l}
|+\rangle\mapsto |++\rangle\\
|-\rangle\mapsto |--\rangle
\end{array}\right.\,.
\]
define the same linear map. Now compare the 
matrices of $|0\rangle, |1\rangle, |+\rangle, |-\rangle$ and the `elements'  $z_{0}, 
z_{1},x_{0}$: 

\begin{center}
\begin{tabular}{|c|c|c|}\hline
$\textbf{FdHilb}$&Matrix Rep.&$\textbf{FRel}$\\ \hline
$|0\rangle$ \begin{tabular}{c} classical for Z\\unbiased for X\end{tabular}&\(\left(\begin{array}{c} 1\\ 0 \end{array}\right)\)&$z_{0} \begin{tabular}{c} classical for Z\\unbiased for X\end{tabular}$\\ \hline
$|1\rangle$ \begin{tabular}{c} classical for Z\\unbiased for X\end{tabular}&\(\left(\begin{array}{c} 0\\ 1 \end{array}\right)\)&$z_{1} \begin{tabular}{c} classical for Z\\unbiased for X\end{tabular}$\\ \hline
$|+\rangle$ \begin{tabular}{c} classical for X\\unbiased for Z\end{tabular}&\(\left(\begin{array}{c} 1\\ 1 \end{array}\right)\)&$x_{0} \begin{tabular}{c} classical for X\\unbiased for Z\end{tabular}$\\ \hline
$|-\rangle$ \begin{tabular}{c} classical for X\\unbiased for Z\end{tabular}&\(\left(\begin{array}{c} 1\\ -1 \end{array}\right)\)& none \\ \hline
\end{tabular}
\end{center}
There is no `$x_{1}$' in the bottom right corner, because there is no element in $\mathbb{B}$ to play 
play the role of $-1$. This lack of negatives in $\textbf{FRel}$ is the reason for the strange asymmetry 
between its `qubit' observables.  But it's quite remarkable that while we do not have \em explicit \em negatives, they are \em implicitly \em present within the basis structures, enabling $\textbf{FRel}$ to simulate several quantum-like features. On the other hand, that the $\textbf{FRel}$ qubit only has two complementary observables as compared to the genuine qubit's three, is again down to the deficiency of $\mathbb{B}$ - there is no element which can 
play the role of $i$ and generate phases.  
We will again be able to \em fix \em this situation without `leaving' $\textbf{FRel}$.

\section{\textbf{Spek}}

Our second example is a sub-category of $\textbf{FRel}$, which is able to give a more complete description of QM. 
\begin{definition}
The objects of $\textbf{Spek}$ are sets of the form $\four\times \ldots\times\four$, where $\four=\{1,2,3,4\}$, together with $\one$. We conveniently subject these to the congruence $\four\times\one=\one\times\four=\four$ and assume strictness of associativity. The morphisms of $\textbf{Spek}$ are all relations generated by relational composition, cartesian product of relations, and relational converse from:
\begin{itemize}
   \item all permutations $\{\sigma_i:\four\to\four\}_i$ on four elements\,; 
   \item a (copying) relation $\delta_Z: \four \rightarrow \four\otimes \four$ defined by:
   \[\begin{array}{llll}
   1\sim \{(1,1), (2,2)\}&\quad 2\sim \{(1,2), (2,1)\}&\quad 3\sim \{(3,3), (4,4)\}&\quad 4\sim \{(3,4), (4,3)\}\,;
   \end{array}\]
   \item a (corresponding deleting)  relation $\epsilon_Z: \four \rightarrow \one::\{1,3\}\sim *$\,.
\end{itemize}
\end{definition}
Hence $\textbf{Spek}$ is a sub-$\dagger$-SMC of $\textbf{FRel}$. The relation $\delta_Z$ can be conveniently represented by the diagram 
   \begin{equation}
   \def\JPicScale{0.8}
   \input{spekdelta.pst}
   \end{equation}
which, if $x\sim (y,z)$,  has $x$ in the $(y,z)$-location of the grid.
Which relations inhabit \textbf{Spek}? By applying different permutations to $x_0:=\epsilon_Z^\dagger$  we get six distinct states, i.e.~morphisms of type $\one \rightarrow \four$:
\begin{equation}
\begin{array}{ccccccc}\label{spekstates}
&z_0::  \ast \sim \{1, 2\} &\qquad\qquad& x_0:: \ast \sim \{1, 3\} &\qquad\qquad& y_0:: \ast \sim \{1, 4\}& \\
&z_1::  \ast \sim \{3, 4\} &\qquad\qquad& x_1:: \ast \sim \{2, 4\} &\qquad\qquad& y_1:: \ast \sim \{2, 3\}& 
\end{array}
\end{equation}
The reader might be somewhat confused by setting $x_0$ rather than $z_0$ or $z_1$ equal to $\epsilon_Z^\dagger$.  The reason is that the index $Z$ in $\delta_Z$ points at `the relation which copies the $Z$-basis'. A corresponding deleting operation needs to be unbiased to this $Z$-basis, e.g.~a basis vector of the $X$-basis. One indeed easily verifies that $(\four,\delta_Z,\epsilon_Z)$ is a basis structure, that this basis structure has two classical points, $z_{0}$ and $z_{1}$, and that is has four unbiased points $y_{0}$, $y_{1}$, $x_{0}$ and $x_{1}$.  Furthermore, composing $\delta_Z$ with various permutations yields three further copying operations $\delta_{Z}^{'}$, $\delta_{Z}^{''}$ and $\delta_{Z}^{'''}$, such that $(\four,\delta_{Z}^{'},x_{1}^\dagger), (\four,\delta_{Z}^{''},y_{0}^\dagger), (\four,\delta_{Z}^{'''},y_{1}^\dagger)$ also form basis structures, with the same classical and unbiased points as $(\four,\delta_Z,\epsilon_Z)$. We will refer to this family of four basis structures which share the same classical points as an \em observable\em.\footnote{To understand this freedom in choosing a deleting operation in the context of ${\bf FdHilb}$ we need to pass from vectors to one-dimensional subspaces i.e.~eliminate redundant global phases. Then, this choice corresponds to fixing  \em coherent superpositions\em.  For the above four cases this would mean $|-\rangle+_1|-\rangle:= |-\rangle+|-\rangle$, $|-\rangle+_2|-\rangle:= |-\rangle+ i |-\rangle$, $|-\rangle+_3|-\rangle:= |-\rangle- |-\rangle$ and $|-\rangle+_4|-\rangle:= |-\rangle- i|-\rangle$. The papers \cite{CD} and \cite{Spekkens} both discuss this subtle issue in great detail.}
We set $Z:=\{(\four,\delta_Z,x_{0}^\dagger), (\four,\delta_Z,x_{1}^\dagger), (\four,\delta_Z,y_{0}^\dagger), (\four,\delta_Z,y_{1}^\dagger)\}$.
We can form new observables by transforming given ones with further permutations. Setting $\delta_X := (\sigma_{(23)}\otimes \sigma_{(23)}) \circ \delta_Z \circ \sigma_{(23)}$, which is represented by
\begin{equation} 
\def\JPicScale{0.8}
\input{spekdeltaprime.pst}
\end{equation} 
we obtain a second observable $X:=\{(\four,\delta_X,y_{0}^\dagger), (\four,\delta_{X}^{'},y_{1}^\dagger), (\four,\delta_{X}^{''},z_{0}^\dagger), (\four,\delta_{X}^{'''},z_{1}^\dagger)\}$ with $x_0$ and $x_1$ as its classical points, and setting $\delta_Y := (\sigma_{(24)}\otimes \sigma_{(24)}) \circ \delta_Z \circ \sigma_{(24)}$, represents by
\begin{equation} 
\def\JPicScale{0.8}
\input{spekdeltaprimeprime.pst}
\end{equation} 
we obtain a third observable $Y:=\{(\four,\delta_Y,x_{0}^\dagger), (\four,\delta_{Y}^{'},x_{1}^\dagger), (\four,\delta_{Y}^{''},z_{0}^\dagger), (\four,\delta_{Y}^{'''},z_{1}^\dagger)\}$  with $y_0$ and $y_1$ as its classical points.  Further transformations under permutations yield nothing further!

\begin{definition}
We call two observables  $A$ and $B$ \em complementary \em if  there exist  basis structures $(X,\delta_A,\epsilon_A)\in A$ and  $(X,\delta_B,\epsilon_B)\in B$ which are complementary --either in the sense of  Def.~\ref{def:complementary} or of Thm.~\ref{thm:complementary}.  
\end{definition}

\begin{theorem}
Observables $X,Y,Z$ are all mutually complementary in the sense of both Def.~\ref{def:complementary} and Thm.~\ref{thm:complementary}.
\end{theorem}

Importantly, the relation $\delta_\oplus:\four\to\four\times\four:: i\sim(i,i)$, represented by 
\begin{equation} 
\def\JPicScale{0.8}
\input{notspekdelta.pst}
\end{equation} 
is not included in ${\bf Spek}$!  This means that ${\bf Spek}$ does not inherit the basis structure on $\four$  from ${\bf FRel}$ which arises from the underlying biproduct structure.  In the light of the previous section this means that  ${\bf Spek}$ inherits the `unexpected' basis structures from ${\bf FRel}$. In particular, these all now stand on equal footing, that is, there is no preferred one anymore. 

As discussed above we can now form `Bell states' 
\[
\eta_{\four}:=\delta_Z\circ\epsilon_Z:\one\to\four\times\four::\ast\sim\{(1,1),(2,2),(3,3),(4,4)\}
\] 
which we can be represented as 
\begin{equation}\label{spekent}
\def\JPicScale{0.8}
\input{spekent.pst}
\end{equation}
We can also form $\eta_{\four\times\ldots\times\four}$ in the obvious manner from $\eta_{\four}$ via monoidal structure. 

\begin{corollary}
The category ${\bf Spek}$ is dagger compact closed.
\end{corollary}
\noindent{\bf Proof.} Equational requirements on compact closure are inherited from ${\bf FRel}$.
\endproof

\section{Connection with Spekkens's toy theory}

The category \textbf{Spek} was not arbitrarily named and defined. It is intended to provide a categorical model of Spekkens's toy model of quantum mechanics. The states of \textbf{Spek} are intended to coincide with the \emph{epistemic} states of Spekkens's theory. 

The single system states of the toy theory are easily seen to correspond exactly with the six states of the \textbf{Spek} object $\four$ in prescription (\ref{spekstates}). 

Two-system states in the theory come in two types. Essentially they are all derived from one of these two states by permutation: 
\begin{equation}
\begin{tabular}{ccc} 
\def\JPicScale{0.8}
\input{spekent.pst}& &\def\JPicScale{0.8}\input{speksep.pst}
\end{tabular}
\end{equation}
The first type, which are the toy theory's analogues of entangled states were derived from the \textbf{Spek} generators above in picture (\ref{spekent}). The second type can also be derived by composition of generators: 
\[
\delta_Z\circ z_0=z_0\times z_0:\one\to\four\times\four::\ast \sim \{(1,1),(1,2),(2,1),(2,2)\}
\]
These correspond to disentangled bipartite states.

With three systems a new type of state appears in Spekkens's toy theory. This is the analogue of a \emph{GHZ state}. Once again, this state can be derived from the  generators of the category \textbf{Spek} via partial transposition:
\[
GHZ:=(\delta_Z\times 1_{\four})\circ\eta_{\four}\,.
\]
With this state in place all other three-system states can straightforwardly be generated by permutation, or by Cartesian product of two-system and one-system states. 

\textbf{Spek} also contains the operations on systems allowed by Spekkens's toy theory. For example, in the case of a single system, permutations model unitary evolution, which are all included  by construction.  Projection caused by measurement is modelled by relations in \textbf{Spek} like: 
\[
z_0 \circ z_{0}^{\dagger} : \four \to \four :: \{1,2\} \sim \{1,2\}\,. 
\]
These operations were not explicitly exposed in \cite{Spekkens}; Spekkens only considered \em functions \em on the underlying sets while $z_0 \circ z_{0}^{\dagger}$ is a \em proper relation \em both involving argument without image as well as multi-valuedness. But there is no reason for excluding these `projection relations' according to the principles outlined in \cite{Spekkens} from which the toy theory is produced.  Moreover,  in \cite{SpekkensBis} Spekkens proposed some axioms for his toy theory, which included map-state duality, and hence inclusion of these projection relations is essential, and more generally, relations such as
\[
x_0 \circ z_{0}^{\dagger} : \four \to \four :: \{1,2\} \sim \{1,3\}\,,
\]
as the counterparts to disentangled bipartite states.

In this way can reproduce all the ingredients of Rob Spekkens's toy theory \cite{Spekkens,SpekkensBis}.  Note that we started from nothing more than a \emph{symmetry group}, a \emph{copying map} and a \emph{deleting map}, together with the \emph{principle of compositionality}.

Furthermore, since all of the generators of \textbf{Spek} correspond to valid states or operations in the theory, and since composition, Cartesian product and relational converse all have counterparts in the theory, \textbf{Spek} is essentially the compositional closure of the toy theory, at least up to the case of three systems. 

Thus we can conclude that at least some of the success of Spekkens's toy theory in modelling QM (for example the teleportation and dense coding protocols, the no-cloning and no-broadcasting theorems) can be attributed to the fact that it, like standard Hilbert space QM, is an instance of a $\dagger$-SMC with basis structures. Ongoing work seeks to establish whether all the results following from the toy theory can be accounted for within the categorical framework. 

\section{Bloch sphere picture}

The states of ${\bf Spek}$ can now be used to interpret  the three perpendicular directions on the Bloch sphere, spanned by the $X$-, $Y$ and $Z$-bases respectively:
   \vspace{-3.3cm}\[
   \def\JPicScale{0.5}
   \quad \qquad \input{HilbBloch.pst}\qquad\qquad\input{SpekBloch.pst}
   \]
In contrast, the states of the Bloch sphere captured by ${\bf FRel}$ are:
   \vspace{-3.3cm}\[
   \def\JPicScale{0.5}
    \quad \qquad\input{FRelBloch.pst}
   \]

\section{Acknowledgements}

Yiannis Hadjimichael pointed at  some typos in the previous version of this paper.

\end{document}